\documentstyle[aps,preprint,tighten]{revtex}
\begin {document}
\draft

\title{Occupation probability of harmonic-oscillator quanta for 
microscopic cluster-model wave functions}

\author{Y. Suzuki$^{1}$, K. Arai$^{2}$, Y. Ogawa$^{3}$, and K. Varga$^{1,4}$
\
\
\\ $^{1}$Department of Physics, Niigata University, Niigata 950-21, Japan
\\ $^{2}$Graduate School of Science and Technology, Niigata University, 
Niigata 950-21, Japan
\\$^{3}$ RIKEN, Hirosawa, Wako, Saitama 351-01, Japan 
\\ $^{4}$Institute of Nuclear Research of the Hungarian Academy of 
Sciences,\\ Debrecen, P.O. Box 51, H-4001, Hungary }
%\date{\today}
\maketitle
\begin{abstract}
We present a new and simple method of calculating the occupation probability 
of the number of total harmonic-oscillator quanta for a 
microscopic cluster-model wave function. 
Examples of applications are given to the recent  
calculations including $\alpha+n+n$-model for $^6$He, $\alpha+t+n+n$-model 
for $^9$Li, and $\alpha+\alpha+n$-model for $^9$Be as well as the 
classical calculations of $\alpha+p+n$-model for $^6$Li and 
$\alpha+\alpha+\alpha$-model for $^{12}$C. 
The analysis is found to be useful for 
quantifying the amount of excitations across the major shell as well as 
the degree of clustering. The origin of the antistretching 
effect is discussed.  
\end{abstract}

\vskip 1.cm
\narrowtext
\par\indent
The microscopic cluster model (MCM) is a many-nucleon theory 
which provides a unified picture of bound-state properties of nuclei 
and nuclear 
reactions. (See, for example, \cite{langanke}.) It is based on 
the assumption that the nucleons in the nuclei form 
substructures, called clusters.  Though the MCM 
is capable of describing a variety of 
structure, its application has mostly been limited to two- or 
three-cluster system. Recent advances in the 
MCM have, however, enabled one to 
treat systems containing more than three clusters and thereby give 
a detailed description of light nuclei including halo nuclei 
\cite{VSL,suzuki,SVAO,vst}. This extension of the applicability has been 
made possible by the inclusion of clusters other than the 
$\alpha$-particle and by the use of the stochastic variational method 
\cite{VSL,VS}. The MCM, as a microscopic theory, 
attempts to derive the nuclear properties from a many-body Hamiltonian 
using a fully antisymmetrized wave function. It 
has, therefore, some relationship to the shell model. Efforts  
have been made to relate the wave function of the MCM 
to that of the $SU(3)$ \cite{hecht} or symplectic \cite{symplectic} 
shell model.    
Such efforts are useful because 
one can compare the model space of the shell model and the MCM.
It is hard to analyse a general MCM wave 
function in terms of shell-model configurations. We will  
show instead that it is easy to calculate the percentage of the 
harmonic-oscillator (HO) excitations involved in the MCM wave function.   

The MCM wave function consists of the 
antisymmetrized product of the intrinsic wave functions of the clusters 
and the relative motion functions between the clusters. The intrinsic 
wave functions are usually assumed to be described with simple 
configurations, whereas the relative motion functions are treated 
flexibly enough to accommodate various types of correlations 
between the clusters as well as large spatial extension if necessary. 
The MCM wave function thus may contain a large number of HO quanta.
To link it with the HO shell-model basis, it is 
useful to calculate the occupation probability of the number of total 
HO quanta. The occupation probability $P_Q$ 
of a definite number of total HO quanta $Q$ for $A$-nucleon 
system is obtained by 
calculating the expectation value of the operator $\cal O$
\begin{equation}
{\cal O}=\frac{1}{2\pi}\int_0^{2\pi}d\theta \ {\rm exp}\Bigl(
i\theta\ (\ \sum_{i=1}^A P_i\  
[H_{\rm HO}(i)-\frac{3}{2}]-Q)\Bigr). 
\end{equation}
Here $P_i$ projects out either proton or neutron. 
It is set the unit 
operator when one calculates the number of total quanta occupied by both 
protons and neutrons. $H_{\rm HO}(i)$ is the 3-dimensional 
HO Hamiltonian divided by $\hbar\omega=\frac{2\hbar^2}
{m}\gamma$.

The MCM wave function is conveniently generated from 
the Slater determinants of the Gaussian wave-packet single-particle (sp) 
functions  
\begin{equation}
\phi_{\kappa}({\bf s}_1,...,{\bf s}_A)={\cal A} \Biggl\{ 
\prod_{i=1}^{A}{\hat \varphi}_{{{\bf s}_i}{\sigma_i}{\tau_i}}^{\nu}
({\bf r}_i) \Biggr\}={\cal A} \Biggl\{ 
\prod_{i=1}^{A}\varphi_{{\bf s}_i}^{\nu}({\bf r}_i)
\chi_{{1\over 2}\sigma_i}{\cal X}_{{1\over 2}\tau_i} \Biggr\} ,
\end{equation}
with  
\begin{equation}
\varphi_{\bf s}^{\nu}({\bf r})=\left({2 \nu \over \pi}
\right)^{3/4} {\rm e}^{-\nu ({\bf r}-{\bf s})^2}.  
\end{equation}
Here ${\cal A}$ is the antisymmetrizer and 
$\kappa=(\sigma_1\tau_1,...,\sigma_A\tau_A)$ stands for the set of the 
spin-isospin quantum numbers of the nucleons. The ${\bf s}_i$ 
parameter or ``generator''
coordinate is a variational parameter in the generator coordinate 
method calculations or it is 
used in an integral transformation \cite{VS,Kami} to derive the matrix 
elements between the Gaussian basis functions \cite{Hack}. 

The matrix element of the operator 
$\cal O$ between the Slater determinants is given by  
\begin{equation}
\langle \phi_{\kappa} ({\bf s}_1,...,{\bf s}_A) \vert {\cal O} \vert 
\phi_{\kappa'}({{\bf s}'}_1,...,{{\bf s}'_A}) \rangle 
=\frac{1}{2\pi}\int_0^{2\pi}d\theta \ {\rm exp}(-iQ\theta)\  
{\rm det} \lbrace B \rbrace, 
\end{equation}
where the matrix $B$ is defined by ({\it i, \,j}=1,...,$A$)
\begin{equation}
B_{ij}=
\langle {\hat\varphi}_{{\bf s}_i\sigma_i\tau_i}^{\nu} 
\vert {\rm exp}\Bigl(i\theta P [H_{\rm HO}-\frac{3}{2}]\Bigr) \vert 
{\hat\varphi}_{{{\bf s}'}_j{\sigma}'_j{\tau}'_j}^{\nu} \rangle. 
\end{equation}
Since the constants, $\nu$ and $\gamma$, are in general 
different, the calculation of the sp matrix element 
of Eq. (5) may seem difficult at first sight but can be 
performed rather easily with the use of the following formulae
\begin{eqnarray}
\varphi_{\bf s}^{\nu}({\bf r})=\Bigl(\frac{\nu\gamma^3}{\pi^2(\nu-\gamma)^2}
\Bigr)^{3/4}\int d{\bf t}\  {\rm exp}\Bigl(-\frac{\nu\gamma}{\gamma-\nu}(
{\bf t}-{\bf s})^2\Bigr)\ \varphi_{\bf t}^{\gamma}({\bf r}),\\
{\rm exp}\Bigl(i\theta [H_{\rm HO}-\frac{3}{2}]\Bigr) 
\varphi_{\bf t}^{\gamma}({\bf r})={\rm exp}\Bigl(-\frac{\gamma}{2}(1-z^2)
{\bf t}^2\Bigr)\ \varphi_{z\bf t}^{\gamma}({\bf r}),
\end{eqnarray}
where $z=e^{i\theta}$. One can prove Eq. (7) by noting that the function,  
$\varphi_{\bf t}^{\gamma}({\bf r})$, is the generating function for  
3-dimensional HO functions. Using Eqs. (6) and (7) 
in Eq. (5) yields the needed matrix element 
\begin{eqnarray}
& &\langle {\hat\varphi}_{{\bf s}\sigma\tau}^{\nu} 
\vert {\rm exp}\Bigl(i\theta P [H_{\rm HO}-\frac{3}{2}]\Bigr) \vert 
{\hat\varphi}_{{\bf s}'{\sigma}'{\tau}'}^{\nu} \rangle 
=\Bigl(\frac{4\nu\gamma}{(\nu+\gamma)^2-(\nu-\gamma)^2\bar{z}^2}
\Bigr)^{3/2}\nonumber \\
& &
\times\ {\rm exp}\Biggl(-\nu\gamma\frac{\nu+\gamma+(\nu-\gamma)\bar{z}^2}
{(\nu+\gamma)^2-(\nu-\gamma)^2\bar{z}^2}({\bf s}^2+{\bf s'}^2)+
\frac{4\nu^2\gamma\bar{z}}{(\nu+\gamma)^2-(\nu-\gamma)^2\bar{z}^2}
{\bf s}\cdot{\bf s'}
\Biggr)\delta_{\sigma,\sigma'}\delta_{\tau,\tau'},
\end{eqnarray} 
where $\bar{z}=z$ or 1 in accordance with  
$\langle\tau \vert P \vert\tau\rangle=1$ or 0. 
 
The value of $\nu$ is usually chosen to give an appropriate size 
for the cluster, while the value of $\gamma$ is determined by the size 
of the whole nucleus. Hence the value of $\nu$ is usually larger than that 
of $\gamma$. The integral in Eq. (6) then does not converge. 
However, even in this case Eq. (8) may safely be used because the 
matrix element should be continuous at all values of $\nu$ and $\gamma$ 
and should have a finite value.  We have confirmed the validity of Eq. (8) 
by comparing the occupation probability calculated by two ways for the 
Gaussian wave-packet, $\varphi_{\bf s}^{\nu}$:  
One is to use Eq. (8) and to change the integration  
variable $\theta$ to the complex variable $z$ 
on a unit circle $|z|=1$. The other is to 
calculate directly the overlap between $\varphi_{\bf s}^{\nu}$ 
and 3-dimensional HO functions. Both methods have given the same result.  

The sp matrix element of $H_{\rm HO}$ itself   
\begin{eqnarray}
& &\langle {\hat\varphi}_{{\bf s}\sigma\tau}^{\nu} 
\vert P [H_{\rm HO}-\frac{3}{2}] \vert 
{\hat\varphi}_{{\bf s}'{\sigma}'{\tau}'}^{\nu} \rangle 
=\Bigl(\frac{3(\nu-\gamma)^2}{4\nu \gamma}-\frac{\nu^2-\gamma^2}
{4\gamma}({\bf s}^2+{\bf s'}^2)+\frac{\nu^2+\gamma^2}{2\gamma}{\bf s}
\cdot{\bf s'}\Bigr)\nonumber \\
& &\times{\rm e}^{-\frac{\nu}{2} ({\bf s}-{\bf s'})^2}\ 
\delta_{\sigma,\sigma'}\ \langle\tau \vert P \vert\tau'\rangle 
\end{eqnarray} 
is needed to calculate the average number of total HO quanta 
contained in the wave 
function. Recently this quantity is used in Ref. \cite{kanada}.

The summation in the exponent of Eq. (1) runs over all the nucleons 
and the probability calculated with it in general contains the 
contribution from the center-of-mass (cm) motion unless the wave function 
is free from the spurious cm motion. Though the Slater 
determinants of Eq. (2) contain the dependence on the cm 
variable, our MCM wave functions generated from them by an integral 
transformation 
turn out to be free from any problem with the spurious cm motion 
\cite{VS}. The probability calculated below is thus a purely intrinsic 
quantity. 

A generalization to a combined occupation probability 
is straightforward. For example, the probability $P_{Q_1,Q_2}$ that 
protons have $Q_1$ quanta and neutrons $Q_2$ quanta or spin-up nucleons 
have $Q_1$ quanta and spin-down nucleons $Q_2$ quanta is obtained 
by using Eq. (1) twice and noting the commutability of the 
corresponding operators. 

As an illustrative example let us consider Brink's  $\alpha+\alpha$-model 
for $^8$Be \cite{brink}. The intrinsic wave function of the 
$\alpha$-particle is constructed from the Slater  
determinant of a 0$s$ HO function with a size 
parameter $\nu$. When the two 
$\alpha$-particles are separated by $S$ and their relative orbital 
angular momentum is $L$, $P_Q$ is calculated by
\begin{equation}
P_Q= \frac{1}{2\pi i}\oint_{|z|=1}dz\ \frac{f(z)}{z^{Q+1}},
\end{equation}
with
\begin{eqnarray}
& &f(z)=\frac{1}{{\rm e}^{-2d}(i_{L}(2d)-4i_{L}(d)+3\delta_{L,0})}
\Bigl(\frac{4\rho}{(1+\rho)^2-(1-\rho)^2z^2}\Bigr)^{12} \nonumber 
\\
& &\times\ {\rm exp}\Bigl(-4\rho d \frac{1+\rho+(1-\rho)z^2}
{(1+\rho)^2-(1-\rho)^2z^2}\Bigr)\\
& &\times\Bigl(i_{L}(\frac{8\rho dz}{(1+\rho)^2-(1-\rho)^2z^2})
-4i_{L}(\frac{4\rho dz}{(1+\rho)^2-(1-\rho)^2z^2})+3\delta_{L,0}\Bigr),
\nonumber
\end{eqnarray}
where $\rho=\gamma/\nu$, $d=\nu S^2$, and $i_{L}(z)=\sqrt{\frac{\pi}{2z}}
I_{L+\frac{1}{2}}(z)$ are the Modified Spherical Bessel functions of 
the first kind. Since $f(z)$ is analytic in the unit circle, Cauchy's 
integral formula can be applied in Eq. (10) to yield $P_Q=f^{(Q)}(0)/Q!$. 
The function $f(z)$ has a leading term proportional to $z^{{\rm max}
(L,4)}$ near $z=0$. Hence $P_Q$ vanishes for $Q<{\rm max}(L,4)$, 
which is the consequence of the Pauli principle. Figure 1 shows the 
$P_Q$ values ($Q$=4, 6, and 8) for $L$=0, 2, 4, and 6 as a function of $S$, 
with a choice of $\nu=0.25$ fm$^{-2}$ and $\gamma=0.15$ fm$^{-2}$. 
The $Q$-dependence is not the same within the $L$=0, 2, and 4 
members but different at each $S$ particularly in the interval of 
2$\sim$4 fm, where the energy functional is known to have a minimum. 
A maximum of $P_{Q=4}$ appears around $S$$\sim$2.9 fm in 
case of $L$=0, while it shifts to a smaller separation of $S$$\sim$2.2 fm 
in case of $L$=4. The diagonal energy curve of $^8$Be as a function of $S$ 
is expected to have a local minimum around the point where $P_{Q=4}$ 
reaches a maximum. Then the difference in the $Q$-dependence on 
$L$ and $S$ values accounts for the antistretching \cite{anti} that the 
energy functional of the cluster-model calculation shows a shrinkage 
of the cluster separation when  
$L$ approaches the value of a band termination.  
 
Table I lists the $P_Q$ values in \% for nucleons, protons, and neutrons for 
some of the wave functions obtained in our recent MCM 
calculations \cite{SVAO,vst,be9} using the 
Minnesota potential \cite{minnesota}. A common value of 
$\nu$=0.26 fm$^{-2}$ is used to describe the intrinsic wave functions of 
$\alpha, t$, and $h$ clusters. The choice of $\gamma$ has some influence 
on the probability. It is set 0.17 fm$^{-2}$ ($\hbar\omega=14.4$ MeV), 
a standard value used in a shell-model calculation for $p$-shell 
nuclei. The value is, however, reduced to 0.13 and 0.14 fm$^{-2}$ for 
$^6$He and $^6$Li, respectively, to take into account their significantly 
large sizes. For the sake of 
reference the calculated root-mean-square (rms) radii for nucleons 
(matters), protons, and neutrons are included in the table. The $P_Q$ 
values are given as a function of $Q_{\rm exc}=Q-Q_{\rm min}$, where 
$Q_{\rm min}$ is the minimum number of HO quanta for the 
lowest Pauli-allowed configuration. The lowest 0$\hbar
\omega$ component is around 50-60 \% for most cases and the sum of 
0, 2, and 4$\hbar\omega$ components accumulates to about 90 \%. The 
admixtures of higher components than $Q_{\rm exc}=4$ are significant in 
the ground states of $^6$Li and $^9$Be and even larger in the ground 
state of $^6$He corresponding to its extended halo structure [4(b)]. The 
probability distribution 
spreads out to a very large number of HO quanta in $^8$Be and the  
$0^{+}_{2}$ state of $^{12}$C, well-known cluster states. They are 
described as a bound state in a large basis. Our wave 
functions for $^{12}$C are similar to those of Ref. 
\cite{kamimura}, which reproduces many properties of $^{12}$C in the 
3$\alpha$-model. The parameter of the Minnesota potential is set to 
reproduce the energy of the $0^{+}_{2}$ state. The ground state energy 
becomes then about 4.5 MeV lower than experiment. The calculated 
monopole matrix element is 4.0 fm$^2$, which is in reasonable agreement 
with the experimental value of 5.4$\pm$0.2 fm$^2$. 
It is noted that no component is dominant in the $0^{+}_{2}$ state of 
$^{12}$C.  Of course it would be possible to maximize the probability with 
lower $Q$ by choosing a smaller value of $\gamma$. However, the 
probability distribution would then spread to higher HO  
quanta in the ground state of $^{12}$C. A recent large-basis 
shell-model calculation \cite{nocore} also suggests that the low-lying 
states in light nuclei from $^4$He to $^7$Li have significant admixtures of 
large HO quanta. Since the calculation seems promising, it is 
interesting to see how the $0^{+}_{2}$ state of $^{12}$C is incorporated 
in its framework. It is also noted that the components with odd 
$Q_{\rm exc}$ values for protons or neutrons are generally smaller. 
For example, in $^6$He $P_{Q_{\rm exc}(p)=1,Q_{\rm exc}(n)=1}$ is about 
3 \%, whereas $P_{0,2}$ and $P_{2,0}$ are 9 and 11 \%, respectively, and, 
among the probability of 16 \% for 4$\hbar\omega$ excitations, the 
probability with $Q_{\rm exc}(p)$ ($Q_{\rm exc}(n)$) =1 or 3 is  
less than 4 \%. 

Comparing the results for $^{7-9}$Li, we see that 
the probability for neutrons has a larger change in the isotopes 
than that for protons. The change follows that of the neutron radius, 
which is consistent with the change of the neutron separation energy. 
In fact the nucleus $^8$Li has the smallest neutron separation energy 
among the three. Since the MCM consistently 
predicts the largest neutron rms radius for $^8$Li \cite{tanihata}, 
its average value,  
$<Q_{\rm exc}>$, for neutrons is largest among the three. 
A comparison of our result with that of Ref. \cite{kanada} indicates that 
the latter wave functions, giving generally much smaller $<Q_{\rm exc}>$ 
values, are rather close to simple shell-model configurations; e.g., 
$<Q_{\rm exc}>$ for neutrons in their paper is about 0.1 for $^9$Li and 0.5 
even for $^8$Be. This may not be surprising because they use a 
single Slater determinant of the Gaussian wave-packets sp functions. 

The calculational method developed here has nothing to do with the 
assumption of the existence of clusters and can obviously be applied 
to those 
precise wave functions for few-nucleon system which are obtained with 
a sophisticated technique. For example, it will be interesting to 
analyse some of the wave functions obtained in Ref. \cite{VS}. 

In summary, we have presented a new and simple method of calculating the 
occupation probability of the number of harmonic-oscillator quanta. It has 
been 
applied to the analysis of some of the wave functions obtained in a 
microscopic multicluster-model calculation. The analysis is found to be 
useful for clarifying the difference from the shell-model truncation made on 
the basis of a major shell as well as quantifying the degree of clustering 
or spatial extension. The origin of the antistretching effect is discussed 
from the change of the occupation probability in the rotational band. 
     
\bigskip

One of the authors (Y.S.) thanks Dr. J. P. Draayer and Dr. D. J. Millener 
for their interest which helped motivate the present study while he 
stayed at the Institute for Nuclear Theory, University of Washington, in 
November of 1995. This work was supported by
a Grant-in-Aid for Scientific Research (No. 05243102 and No. 06640381) 
of the Ministry of Education, Science and Culture, Japan. Most of the 
calculations were done with the use of RIKEN's VPP500 computer.

\squeezetable
\begin{table}
\caption{The occupation probability of the number of harmonic-oscillator 
quanta for microscopic 
multicluster-model wave functions. The probabilities for nucleons, 
protons, and neutrons are given in \% in the upper, middle, and lower 
rows, respectively, as a function of oscillator excitations. When the 
probability for 
neutrons is the same as that for protons, only the proton case is shown. 
Asterisk indicates the probability of less than 1 \% and dashed line 
represents vanishing probability. The average number of 
oscillator excitations is given in the column labeled $<Q_{\rm exc}>$. 
The details of the wave functions are referred to Ref. [4(b)] for $^6$He 
and $^6$Li, to Ref. [5] for $^7$Li, $^8$Li, $^9$Li, and $^9$C, and to 
Ref. [14] for $^9$Be. }

\vspace {0.5mm}

\begin{tabular}{cccccccccccccccccc}

state & {\rm rms radius}  &  \multicolumn{15}{c}{$Q_{\rm exc}$} & \\
\cline{3-17}

(model) & [fm] &
 0 & 1 & 2 & 3 & 4 & 5 & 6 & 7 & 8 & 9 &10 & 11 & 12 & 13 & 14 &
\raisebox{1.5ex}[0pt]{$<Q_{\rm exc}>$}  \\
\hline

$^6$He(0$^+$) & 
  $r_m=2.51$ & 46 & --- & 23 & --- & 16 & --- & 7  & --- & 4 
& --- & 2 & ---& 1 & --- & $\ast$ & 2.5 \\
($\alpha$+$n$+$n$) & $r_p\;=1.87$ & 63 & 9  & 19 & 3  & 4  & $\ast$ & $\ast$ 
& $\ast$ & $\ast$ & $\ast$ & $\ast$ & $\ast$ & $\ast$ & $\ast$ & $\ast$ &
 0.8  \\
& $r_n\;=2.78$ & 59 & 4  & 14 & 5  & 8  & 2 & 3 & 1 & 1 & 
$\ast$ & $\ast$ & $\ast$ & $\ast$ & $\ast$ & $\ast$ & 1.7   \\
\hline

$^6$Li(1$^+$) & 
  $r_m=2.44$ & 52 & --- & 23 & --- & 13  & --- & 6 & --- & 3 & --- &
2 & --- & $\ast$ & --- & $\ast$ & 2.1 \\
($\alpha$+$p$+$n$) & $r_p\;=2.44$ & 64 & 8  & 15 & 4  &  5  & 2  & 1 
& $\ast$ & $\ast$ & $\ast$ & $\ast$ & $\ast$ & $\ast$ & $\ast$ & $\ast$ &
 1.1   \\
\hline

$^7$Li(3/2$^-$) & 
  $r_m=2.34$ & 63 & --- & 20 & --- & 9  & --- & 4 & --- & 2 & --- &
$\ast$ & --- & $\ast$ & --- & $\ast$ & 1.4  \\
($\alpha$+$t$) & $r_p\;=2.28$ & 77 & 2  & 16 & $\ast$ & 4 
& $\ast$ & $\ast$ & $\ast$ & $\ast$ & $\ast$ & $\ast$ &
$\ast$ & $\ast$ & $\ast$ & $\ast$ & 0.6  \\
& $r_n\;=2.38$ & 73 & 1  & 17 & $\ast$ & 5  & $\ast$ & 1 & $\ast$ 
& $\ast$ & $\ast$ & $\ast$ & $\ast$ & $\ast$ & $\ast$ & $\ast$ & 0.8  \\
\hline

$^8$Li(2$^+$) & 
  $r_m=2.45$ & 61 & --- & 18 & --- & 11  & --- & 4 & --- & 2 & --- & 1 & 
--- & $\ast$ & --- & $\ast$ & 1.7  \\
($\alpha$+$t$+$n$)& $r_p\;=2.19$ & 79 & 6  & 11 & 1  & 2  & 
$\ast$ & $\ast$ & $\ast$  & $\ast$ & $\ast$ & $\ast$ & $\ast$ &
$\ast$ & $\ast$ & $\ast$ & 0.4  \\
& $r_n\;=2.60$ & 67 & 3  & 14 & 2  & 7   & 1 & 2 & $\ast$ & 1 & $\ast$
& $\ast$ & $\ast$ & $\ast$ & $\ast$ & $\ast$ & 1.3  \\
\hline

$^8$Be(0$^+$) & 
 $r_m=3.27$ & 36 & --- & 18 & --- & 12  & --- & 7 & --- & 5 & --- &
4 & --- & 3 & --- & 2 & 7.6  \\
($\alpha$+$\alpha$) & $r_p\;=3.27$ & 47 & --- & 21 & --- & 11 & --- & 6 & 
--- & 4 & --- & 3 & --- & 2 & --- & 1 & 3.8  \\
\hline

$^9$Li(3/2$^-$) & 
  $r_m=2.40$ & 66 & --- & 17 & --- & 11  & --- & 4 & --- & 2 & --- & $\ast$
& --- & $\ast$ & --- & $\ast$ & 1.3  \\
($\alpha$+$t$+$n$+$n$) & $r_p\;=2.10$ & 82 & 6  & 9 & 1  & 1 & $\ast$ 
& $\ast$ & $\ast$ & $\ast$ & $\ast$ & $\ast$ & $\ast$ & $\ast$ & $\ast$
& $\ast$ & 0.4  \\
& $r_n\;=2.54$ & 71 & 3  & 12 & 2  & 6   & 1 & 2 & $\ast$ & $\ast$ 
& $\ast$ & $\ast$ & $\ast$ & $\ast$ & $\ast$ & $\ast$ & 1.0  \\
\hline

$^9$C(3/2$^-$) & 
  $r_m=2.52$ & 60 & --- & 17 & --- & 12  & --- & 5 & --- & 3 & --- & 1
& --- & $\ast$ & --- & $\ast$ & 1.8  \\
($\alpha$+$h$+$p$+$p$) & $r_p\;=2.68$ & 65 & 4  & 12 & 3  & 7 & 1 
& 2 & $\ast$ & 1 & $\ast$ & $\ast$ & $\ast$ & $\ast$ & $\ast$
& $\ast$ & 1.4  \\
& $r_n\;=2.16$ & 79 & 8  & 9 & 2  & 1 & $\ast$ & $\ast$ & $\ast$ & $\ast$ 
& $\ast$ & $\ast$ & $\ast$ & $\ast$ & $\ast$ & $\ast$ & 0.4  \\
\hline

$^9$Be(3/2$^-$) & 
  $r_m=2.50$ & 54 & --- & 21 & --- & 12  & --- & 5 & --- & 3 & --- &
2 & --- & $\ast$ & --- & $\ast$ & 2.1  \\
($\alpha$+$\alpha$+$n$) & $r_p\;=2.39$ & 71 & 3  & 17 & 1  & 5 & $\ast$ & 1 
& $\ast$ & $\ast$ & $\ast$ & $\ast$ & $\ast$ & $\ast$ & $\ast$ & 
$\ast$ & 0.8  \\
& $r_n\;=2.58$ & 65 & 2  & 18 & 1  & 8 & $\ast$ & 3 & $\ast$ & 1 
& $\ast$ & $\ast$ & $\ast$ & $\ast$ & $\ast$ & $\ast$ & 1.3  \\
\hline

$^{12}$C(0$_1^+$) & 
  $r_m=2.20$ & 54 & --- & 30 & --- & 11  & --- & 3 & --- & $\ast$ & --- &
$\ast$ & --- & $\ast$ & --- & $\ast$ & 1.4 \\
($\alpha$+$\alpha$+$\alpha$) & $r_p\;=2.20$ & 70 & 5  & 19 & 1  & 4 & $\ast$ & $\ast$ 
& $\ast$ & $\ast$ & $\ast$ & $\ast$ & $\ast$ & $\ast$ & $\ast$ & $\ast$ 
& 0.7  \\
\hline

$^{12}$C(0$_2^+$) & 
  $r_m=3.75$  & $\ast$ & --- & 11 & --- & 12  & --- & 12 & --- & 10 
& --- & 8 & --- & 7 & --- & 6 & 16.1  \\
($\alpha$+$\alpha$+$\alpha$) & $r_p\;=3.75$  &  5 & 7  & 15 & 7  & 11 & 6 & 8 & 5 & 6 
& 4 & 4 & 3 & 3 & 2 & 2 & 8.2  \\
\end{tabular}
\end{table}

\newpage
\begin{center}
{\large Figure Captions}
\end{center}

\bigskip

\begin{enumerate}

\item[Fig.\,1.] The occupation probability of the number of oscillator 
quanta $Q$ for the $L=0-6$ states of $^8$Be. The wave functions of $^8$Be 
are assumed to be given by Brink's 2$\alpha$-model [12] with a mean 
separation $S$. 
\end{enumerate}
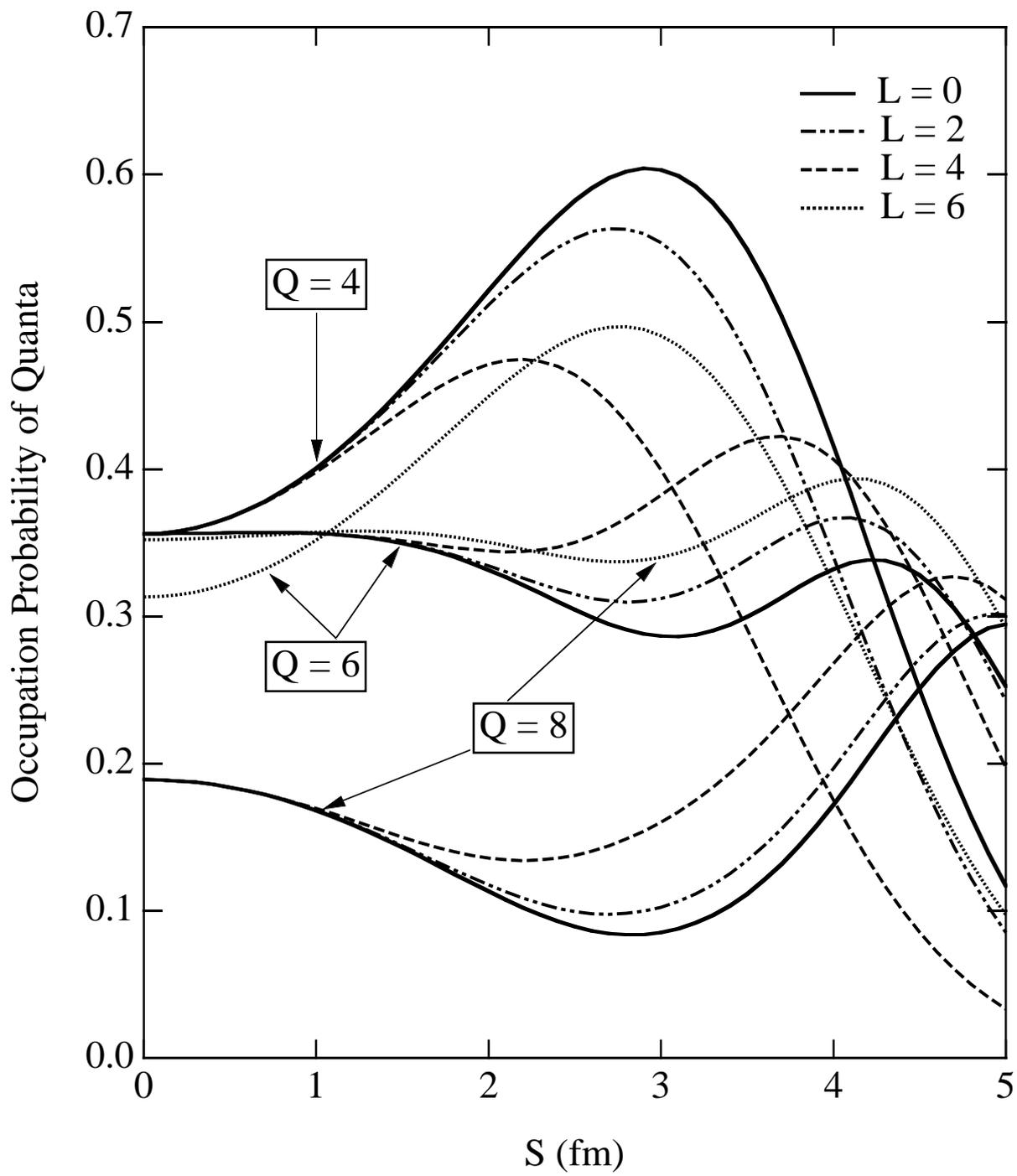
\end{document}